\documentclass[twocolumn]{el-author}
\graphicspath{{/.}}
\DeclareGraphicsExtensions{.eps,.pdf,.jpeg,.png}

\usepackage{cite} 
\usepackage{graphicx} 
\usepackage{subfigure}
\usepackage{url}
\usepackage{amsmath}
\usepackage{tikz}
\usepackage{pgf} 

\usepackage[amssymb]{SIunits}
\usepackage{algpseudocode}
\usepackage{algorithm}
\usepackage{color}
\usepackage{enumitem}

\newcommand{\ua}{\uparrow}
\newcommand{\nc}{\newcommand}
\nc{\da}{\downarrow} \nc{\hc}{\hat{c}} \nc{\hS}{\hat{S}}
\nc{\bra}{\langle} \nc{\ket}{\rangle} \nc{\eq}{equation (\ref}
\nc{\h}{\hat} \nc{\hT}{\h{T}}\nc{\be}{\begin{eqnarray}}
\nc{\ee}{\end{eqnarray}}\nc{\rd}{\textrm{d}}\nc{\e}{eqnarray}\nc{\hR}{\hat{R}}\nc{\Tr}{\mathrm{Tr}}
\nc{\tS}{\tilde{S}}\nc{\tr}{\mathrm{tr}}\nc{\8}{\infty}\nc{\lgs}{\bra\ua,\phi|}\nc{\rgs}{|\ua,\phi\ket}
\nc{\hU}{\hat{U}}\nc{\lfs}{\bra\phi|}\nc{\rfs}{|\phi\ket}\nc{\hZ}{\hat{Z}}\nc{\hd}{\hat{d}}\nc{\mD}{\mathcal{D}}
\nc{\bd}{\bar{d}}\nc{\bc}{\bar{c}}\nc{\mc}{\mathcal}\nc{\ea}{eqnarray}\nc{\mG}{\mathcal{G}}\nc{\bce}{\begin{center}}
\nc{\ece}{\end{center}}

\date{24th Dec 2013}

\begin{document}

\title{Technical Report: A New Fast Multi-Device Wireless Power Transfer Scheme Using an Intermediate Energy Storage}

\author{C.S. Yoon, S.S. Nam, and S.H. Cho}
\abstract{
A new multi-device wireless power transfer scheme that reduces the overall charging time is presented. The proposed scheme employs the intermediated energy storage (IES) circuit which consists of a constant power driving circuit and a super-capacitor. By utilizing the characteristic of high power density of the super-capacitor, the receiver can receive and store the energy in short duration and supply to the battery for long time. This enables the overlap of charging duration between all receivers. As a result, the overall charging time can be reduced.
}
\maketitle

\section{Introduction} \label{sec:1}
As the number of mobile devices increases, the ability to simultaneously charge multiple devices is strongly demanded in wireless power transfer (WPT) system. 
One of possible ways for charging multiple devices is to allocate a dedicated transmitting circuit to each receiver\cite{WPT_2011}. This method provides efficient and stable WPT however requires the same number of transmitting circuits as receivers which are desired to charge simultaneously.
Another method is to employ a power control circuit in each of the receivers\cite{Boys_2000,Kurs_2007}. By using independent power control circuit, the receivers can receive the required power from the constantly transmitted power signal of a single transmitter. This method has the benefit of charging multiple receivers with a single transmitter. However, the transfer efficiency can be degraded.

Recently, the time-division multiplex wireless power transfer (TDM-WPT) scheme was proposed in \cite{Qualcomm_2010}.
This scheme enables wireless power transfer (WPT) to multiple receivers with a single transmitter by using a time-multiplexing technique.
This scheme provides high power transfer efficiency as in \cite{WPT_2011} because accurate controls of the selected receiver is possible during the dedicated time slot. 
In addition, the cost of the system can be reduced as in \cite{Boys_2000,Kurs_2007} by including only a single power driving circuit.
However, for $N$ receivers, the overall charge time becomes $N$ times as long as that of a single receiver.

One of the ways to reduce the charge time is to overlap the durations for charging the batteries of different receivers as much as possible. This time overlapping processes can be achieved by supplying energy to the batteries using temporary energy storage even while the energy cannot be transferred directly from the transmitter. Thus, the batteries of the different receivers can be simultaneously charged. As a result, the total charge time can be reduced by the sum of these overlapped times.

For practical implementation of this conceptual solution, we propose a new TDM-WPT scheme using the intermediate energy storage (IES), which includes a constant power driver circuit and a super-capacitor. Specifically, while the receiver drains power directly from the transmitter, the remaining power after providing the required power to battery can be supplied to the super-capacitor. Because the super-capacitor has the high power density, the IES can receives a relatively wide range of power compared to the battery. Therefore, while the other receivers are directly supplied by the transmitter, the battery of a receiver can be charged indirectly by using the stored energy in the IES.

The remainder of this paper is organized as follows. In section II, we present the detail structure and the simple mode of operation for the proposed system briefly. Section III presents simple numerical analysis of the performance of the proposed system. In particular, we present the overall charge time of the proposed system for the various system parameters. Section IV presents the simulation results based on the assumptions in the numerical analysis. In section V, the simulation considering practical battery charger, which of required power changes continuously, is undertaken.

\section{Structure and operation of IES assisted TDM-WPT scheme} \label{sec:2}
\begin{figure}[!t]
\centering{\includegraphics[width=90mm, clip]{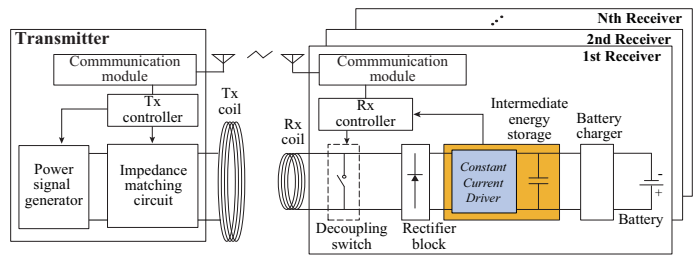}}
\caption{The functional block diagram of the proposed system.}
\label{fig_md_wpt}
\end{figure}

Fig. \ref{fig_md_wpt} shows the proposed system including a single transmitter and multiple receivers. 
Specifically, the system basically contains basic parts such as power conversion part, power transfer part, energy storage part, communication part, and control part, which is previously included in the conventional TDM-WPT system, which described in detail at \cite{Qualcomm_2010}, and an extra simple IES circuit. 
The IES is employed between the rectifier block and the charging circuit.

The power conversion part includes a power signal generator and a rectifier block. the power signal generator produces AC signal of the desired frequency/amplitude  from DC power source and the rectifier block restores DC power from received AC signals. The power transfer part includes Tx/Rx coils and a matching circuits. the coils delivers power through magnetic coupling and the matching circuit controls impedance to provide high transfer efficiency. The energy storing part includes a battery and a battery charger to supply sufficient power to the battery. The communication part provides interfaces and methods for delivering commands and information between a transmitter and receivers. The control part including a Tx/Rx controller provides the overall control of WPT process based on the input which is received through the communication part.

The structure of the IES consists of a constant power driver and a super-capacitor.
The power driver drains constant power from the rectifier block and supplies it to both the super-capacitor and the charging circuit.
The super-capacitor is used for temporary energy storage for supplying the battery charging circuit when the receiver directly cannot supply energy to the charger.

The basic concept to reduce the charge time is to charge multiple receivers in a virtually simultaneous manner. Specifically, the IES and the battery are supplied concurrently when the receiver directly receives the power from the transmitter. When the IES is fully charged, the transmitter selects the next receiver and transfers power. At the same time, in the previous receiver, the IES supplies energy to the battery until the IES is fully discharged in Fig.\ref{IES_time}. As a result, the batteries of multiple receivers can be charged simultaneously.
The detailed modes of operation are listed as follows:

The detailed mode of operations for the proposed system is listed as follows: 
\begin{enumerate}[leftmargin=15pt, topsep=0cm, itemsep=0cm, parsep = 0cm, partopsep = 0cm]
\item[1)] At the initial stage, the transmitter collects the system parameters of all nearby receivers by using the communication unit. Based on the received general informations, the transmitter searches for the optimal parameters of each receivers which provides high transfer efficiency in WPT operation.
\item[2)] Subsequently, the transmitter selects the receiver according to the predetermined order and broadcasts a command message denoting the start of receiving for the selected receiver.
\item[3)] When receivers identify the message, all the receivers control the coupling status to transfer power to the selected receiver. Specifically, the unselected receivers are decoupled immediately. On the other hand, the selected receiver are coupled after the included IES becomes fully discharged. 
\item[4)] As the receiver drains power, the included IES is fully charged. In this point of time, the receiver is decoupled so that the transmitter identifies the termination of receiving power.
\item[5)] The charging processes are continued from 2) until all receivers are fully charged.
\end{enumerate}
the performance of the proposed system will be discussed in the following section.

By employing an IES in each receiver, the following benefits can be achieved additionally. First, the optimal parameter searching processes in 1) is not continuously required unless the locations of the receivers are changed. Because the receiver keeps the fixed load impedance due to the constant power driver of the IES. Second, IES can relieve the receiving power control error by storing the rest of supplied energy which is delivered more than the required. It will help to transfer power with high efficiency by reducing the dissipation of power into heat at the receivers.

\begin{figure}[!t]
\centering
\includegraphics[width=90mm, trim=0cm 0cm 0cm 1cm, clip=true]{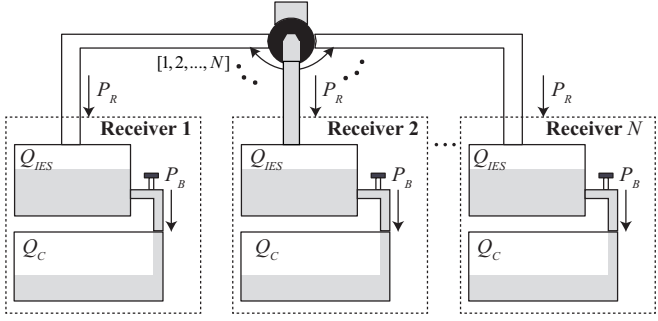}
\caption{The water tank analogy of the proposed TDM-WPT system}
\label{water_tank_model}
\end{figure}

\section{Performance analysis of IES assisted MD-WPT } \label{sec:3}
In the analysis, we assume $N$ identical receivers, each of which is fully discharged at the initial stage.   
The receiver has the following parameters: i) the required power for the battery over the state of charge $R$, namely, $P_B[R]$ ; ii) the receiving power $P_R$ ; iii) the capacity of the battery $Q_C$ ; iv) the capacity of the IES $Q_{IES}$; and v) the required time to switch between receivers $T_d$. 
The above-defined parameters stick to the following conditions: ${P_B} < {P_R}$, and ${Q_{IES}} < {Q_C}$.
Further, in this paper, for tractable analysis, we assume $P_B[t]$ has a constant value $P_B$ over every $t$.

Fig. \ref{water_tank_model} shows the water tank analogy of the proposed system. In this analogy, the water and its transfer rate denote energy and power, respectively.
The central supplying pipe denotes a transmitter and sets of two tanks which is enclosed with the dotted line denote receivers.
Among this set of two tanks, the upper and the lower tanks denote the IES and the target battery, respectively. Each of capacities can be denoted as $Q_IES$ and $Q_C$.
The water, which are supplied through the central pipe, is transferred to the selected site at the rate of $P_R$. And then, this water is supplied to the lower tank at the rate of $P_B$ and stored in the upper tank at the rate of $P_R-P_B$ simultaneously.
Even while water cannot supply to the lower tank directly, the upper tanks extend the supplement time by using the stored water.
 
Based on the proposed system, a number of cycles for charging and discharging the IES are required to completely charge a single receiver because of the capacity constraints of the IES. 
For analytical convenience, in this paper, we consider that cycles of each receiver start from every charging of the IES, which is shown visually in Fig.\ref{IES_time}. 

Specifically, a single cycle includes a charge phase and a discharge phase. 
In the charge phase, from the received power $P_R$, the required power $P_B$ is supplied to the battery, and the remaining power $P_R-P_B$ is supplied to the IES until the IES is fully charged.
In the following discharge phase, the IES supplies energy to the battery at the rate of $P_B$ until the IES is fully discharged.
Accordingly, the durations of the charge and discharge phase can be estimated as $A=\frac{Q_{IES}}{P_R-P_B}$ and $B=\frac{Q_{IES}}{P_B}$, respectively. During both phases, the battery is actually charged, thus the duration for charging battery per cycle can be derived as $C = A + B$.

\begin{figure}[b!]\centering{\includegraphics[width=90mm, clip]{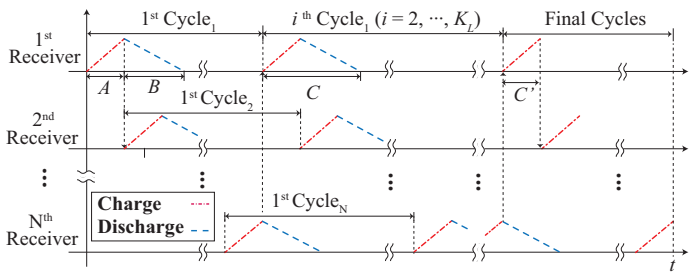}}
\caption{The energy in the IES of $i$-th receivers without switching delay ($i=1,\cdots,N$).}
\label{IES_time}
\end{figure}

For each cycle, the same amount of energy is delivered to the battery and this cycle is repeated before the final cycle.
Thus, the required numbers of repetitive cycles can be estimated as the quotient of the battery capacity divided by the energy transferred to the battery through a cycle, i.e., ${K_L} =  \left\lfloor \frac{{{Q_C}}}{{{P_B}C}} \right\rfloor$, where $\left\lfloor(\cdot)\right\rfloor$ denotes the largest integer not greater than $(\cdot)$.
Then, at the final cycle, the remaining energy is delivered to complete the battery charging process.
Therefore, the required time can be estimated as $C'=\frac{Q_C-K_L P_B C}{P_B}$. 
For any given number of receivers, $N$, each of the receivers requires the same $K_L$ cycles because the receiver delivers the same energy to the batteries per cycle for all $K_L$ cycles, being independent of $N$.

However, when $N$ is larger than the threshold value which will be discussed in detail below, the cycle duration increases.
In this case, the battery charging operation may be temporarily interrupted because the energy stored in the IES is exhausted before the next charge phase.
Here, we define this period of interruption as the standby time. We can then separately consider two cases as follows: case 1 does not have a standby time, and case 2 has a standby time.
As follows, for simple approach, we start the analysis from the cases that the switching time $T_d = 0$.

In case 1, the charge phase of each receiver must return before the IES is fully discharged.  
Thus, the charge phases of the other receivers have to be completed before the discharge phase of a receiver ends, i.e., $(N-1)A \leq B$. 
Note that for given receiver parameters, the maximum $N$ value, which does not have the standby times, can be estimated as
\begin{equation}
N_{max}=\left\lfloor(A+B)/A\right\rfloor.
\label{ST_Bound}
\end{equation}
For this condition, In this condition, the cycle of the charge and discharge phases repeats without the standby times between. Thus, the cycle duration can be estimated as the sum of the durations for a charge phase and a discharge phase, i.e., $C$. On the other hand, the final cycle has a different duration compared to the previous cycles as noted above. Because the start of the final cycle for receivers delayed by $A$ sequentially, the duration for the final cycles of the receivers can be estimated as $(N-1)(A)+C'$.
Thus, the result of $T_{OC}$ can be summarized as
\begin{equation}
\begin{split}
{T_{OC}}&= K_L C + C' + \left(N-1\right)A\\
	&=\frac{Q_C}{P_B}+\left(N-1\right)\left(\frac{Q_{IES}}{P_R-P_B}\right).
\end{split}
\label{T_OC_NO_ST}
\end{equation}

In case 2, the operation ranges of system parameters are inverse of those in case 1, i.e., $(N-1)A < B$. In this case, the duration of a cycle becomes increased to $NA$ because the charge phases of the other receivers must be continued after the discharge phase of a receiver ends. The duration of completing the final cycles of all receivers can have numerous results according to the conditions, which are caused by the remaining energy at the final cycle and charging patterns as followings:
\begin{enumerate}[leftmargin=15pt, topsep=0cm, itemsep=0cm, parsep = 0cm, partopsep = 0cm]
  \item[a)] For small remaining energy, i.e., $C' \leq B/(N-1)$; the duration of charge phase in the final cycle can be decreased. Accordingly, the final cycles become to have the same condition of case 1, thus, each final cycle does not contain any standby time. The $N$th receiver, which starts the discharge phase when the first receiver starts the charge phase of the final cycle, also starts the charge phase without the standby time. Therefore, the time duration can be estimated as $B+C'$.    
  \item[b)] For larger remaining energy than case a, i.e., $B/(N-1) < C' \leq A$; the charge phase duration in the final cycle becomes increased. For this reason, a set of the final cycles have the same condition of case 2, thus, each final cycle contains standby time. In this case, the receivers are switched by interval of the charge phase duration of the final cycle. As a result, the time duration can be estimated as $NC'$
  \item[c)] For larger remaining energy than case b, i.e., $A < C'$; the discharge phase appears in the final cycle of all receivers. In this case, the receivers are switched by the charge phase duration of the full cycle, $A$. Therefore, the time duration can be estimated as $\left(N-1\right)A+C'$.
\end{enumerate}
      
The summarized result of $\underset{T_d=0}{T_{OC}}$ can be expressed as
\begin{equation}
\underset{T_d=0}{T_{OC}} =
\begin{cases}
K_L N A + B + C',			&C' \leq \frac{B}{N-1}\\
K_L N A + N C',				&\frac{B}{N-1}< C' \leq A\\
K_L N A + \left(N-1\right) A + C',	&A < C'  
\label{T_OC_ST} 
\end{cases}
\end{equation}

Eqs. (\ref{T_OC_NO_ST}) and (\ref{T_OC_ST}) are increasing functions of $Q_{IES}$, which means that the overall charging time is minimized by selecting the capacity of the IES as low as possible. However, when the IES has low capacity, the number of switching, which is required to completely charge receivers, considerably increases. Here, by considering a switching time $T_d$, the sum of switching times affects significantly on the overall charge time $T_{OC}$. As a result, there exists an optimal IES capacity $Q_{IES}$ that can minimize $T_{OC}$.

\begin{figure}[t!]\centering{\includegraphics[width=90mm, clip]{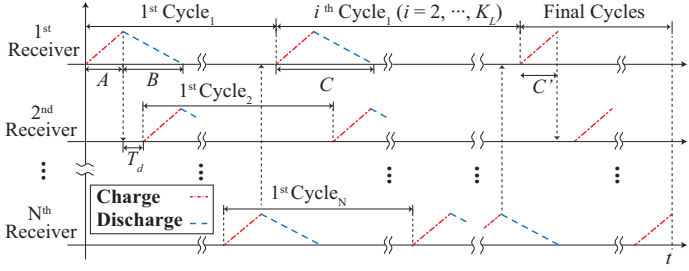}}
\caption{The energy in the IES of $i$-th receivers with switching delay ($i=1,\cdots,N$).}
\label{IES_time}
\end{figure}

By considering the switching delay $T_d$, the duration for charging the IESs of the other receivers increases while the duration for discharging the IES of a receiver remains unchanged.
For this reason, the standby times can exist for same parameters of the condition in case 1.
Therefore, the condition for case 1 has to be updated to $B \leq (N-1)A + N T_d$ by taking account of $N$ operations of switching receivers in a cycle. 
In this updated case, the duration of a cycle is unchanged because the standby time also does not exist as before.
The duration of the last cycle increases by $(N-1)T_d$ because the overlapping interval increases by $T_d$. The summarized result of $T_{OC}$ can be expressed as
\begin{equation}
\underset{T_d>0}{T_{OC}} =(N-1)(A+T_d)+\frac{Q_C}{P_B}
\label{T_OC_NoST_Td_Full}
\end{equation}

The condition for the case 2 is updated to $B > (N-1)A+NT_d$, which is the inverse condition of the updated case 1. In this case, the cycle duration is updated to $N(A+T_d)$ by considering every switching delays between charge phases. 
The duration for final cycles is updated also. For case a), the condition is updated to $(N-1)C'+NT_d \leq B$ because the switching time for the N receivers is considered. The duration decreases by $T_d$ because the discharge time of the first receiver, i.e., $B$ decreases by $T_d$. For condition b), $(N-1)C'+NT_d \leq B, C' < A$; the duration increases by $(N-1) T_d$ because the switching time among the $N$ receivers is considered. For condition c), the duration increases similar to that for condition b) for the same reason. The summarized result of $T_{OC}$ can be expressed as
\begin{equation}
\begin{split}
\!\!\!&\underset{T_d>0}{T_{OC}}\!\!=\!\!
\begin{cases}
K_L\!N\!(\!A\!+\!T_d\!)\!+\!(\!B\!\!-\!T_d\!)\!+\!C',\!&\!C'\leq\frac{\!B\!-\!N\!T_d\!}{(\!N\!\!-\!1\!)}\\
K_L\!N\!(\!A\!+\!T_d\!)\!+\!(\!N\!-\!1\!)\!T_d\!+\!N\!C',\!&\!\frac{\!B\!-\!N\!T_d\!}{(\!N\!\!-\!1\!)}<C'\leq A\\
K_L\!N\!(\!A\!+\!T_d\!)\!+\!(\!N\!-\!1\!)\!(A\!+\!T_d\!)\!+\!C',&A<C'
\end{cases},
\label{T_OC_ST_Td_Full}
\end{split}
\end{equation}

For providing the simple guideline of the practical implementation, the upper bound of the charging time is presented.
The final cycles for all of the receivers can be completed within the sum of the durations of the final cycle and the previous cycle because the final cycle of the $N$ th receiver starts within one cycle duration after the final cycle of the first receiver starts. The error between the actual time and the estimate is less than a single cycle duration.
Thus, the upper bound of $T_{OC}$ not to have the standby time can be summarized as
\begin{equation}
\begin{split}
T_{OC} < &K_L C + C' + C \\ 
&= \left(\frac{Q_C}{P_B} + 1\right) \frac{Q_{IES} P_R}{P_B \left(P_R-P_B\right)}.
\end{split}
\label{T_OC_NoST_Td_Upper}
\end{equation}
the upper bound of $T_{OC}$ to have the standby time can be summarized as
\begin{equation}
\begin{split}
T_{OC}\!< &\!N\!\left(\!A\!+\!T_d\!\right)\!\left(\!\frac{Q_C}{Q_{IES}}\!+\!1\!\right)\\
= &\!N\!\left(\!\frac{Q_{IES}}{P_R\!-\!P_B}\!+\!T_d\!\right)\!\left(\!\frac{Q_C}{Q_{IES}}\!+\!1\!\right).
\end{split}
\label{T_OC_ST_Td_Upper}
\end{equation}

The curves in Eqs. \ref{T_OC_NoST_Td_Upper} and \ref{T_OC_ST_Td_Upper} show convexity over the $Q_{IES}$ value. The curves have an optimal value for a specific $Q_{IES}$ value.
Therefore, the optimization problem is equivalent with the following problem:
\begin{equation} 
\begin{split} 
&\underset{Q_{IES}}{\text{minimize}} \quad \underset{T_d>0}{T_{OC}}(P_R,P_B,Q_{IES},Q_{C},T_d)\\
&\text{subject to} \quad 
\begin{array}{ll}
&Q_{IES} <Q_{C}, \\ 
&P_B < P_R\\
&0 < T_d
\end{array}
\end{split} 
\end{equation}

\section{Simulations} 

MATLAB simulations have been undertaken to evaluate the performance of the proposed scheme and to provide the simple guideline for key parameters. The simulation is conducted based on the assumptions that is explained in the previous section, and it presents the upper bound and actual value of the charge time.
For practical consideration, we assume the parameters to be $P_B = 1 \watt$, $P_R = 4.2 \watt$, $Q_C = 1 \watt\hour$, and $T_d = 1 \milli\second$. 

\begin{figure}[t!]
\centering{\includegraphics[width=8.5cm, trim = 40mm 0mm 40mm 0mm, clip]{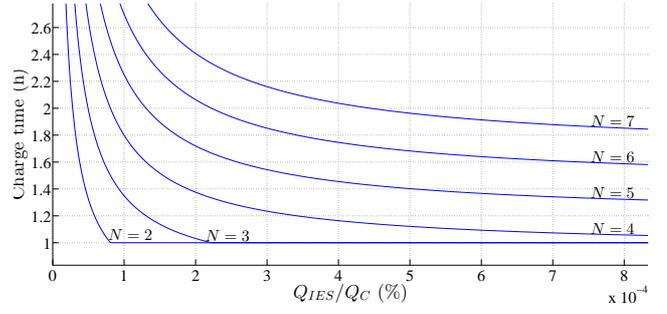}}
\caption{The total charge time over $Q_{IES}$.}
\label{sim_ch_Q_IES}
\end{figure}

Fig. \ref{sim_ch_Q_IES} shows the upper bound of the charge time of the proposed scheme over the capacity ratio of the IES to the battery for the various number of receivers. For each $N$ value, the result has the convex waveform according to (\ref{T_OC_NoST_Td_Upper}) and (\ref{T_OC_ST_Td_Upper}). 
Therefore, the optimal $Q_{IES}$ value exists where $T_{OC}$ is minimized.
However, the reduction of $T_{OC}$ caused by the increase of $Q_{IES}$ decreases as $Q_{IES}$ becomes closer to the optimal point.
Consequently, for the benefit of the user, $Q_{IES}$ can be selected to achieve the required constraints such as charge time, cost, volume, etc.

\begin{figure}[b!]
\centering{\includegraphics[width=9cm, trim = 20mm 0mm 20mm 0mm, clip]{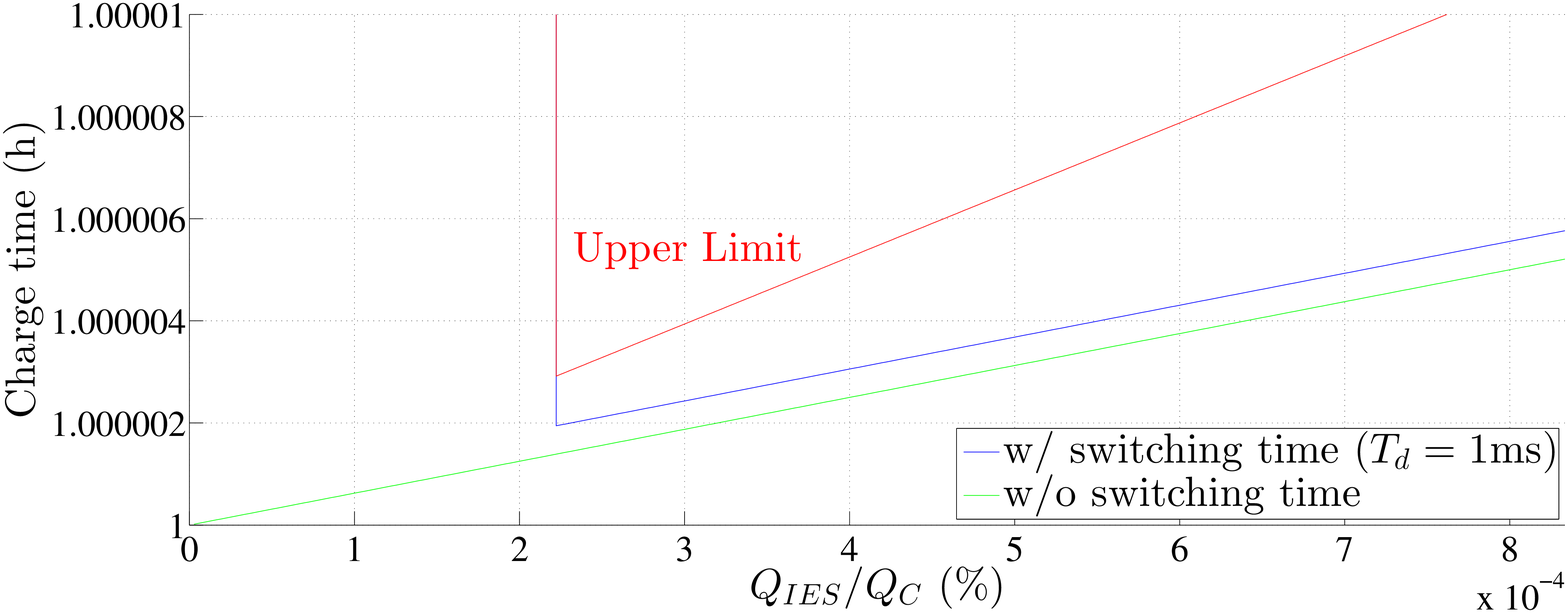}}
\caption{The charging time over $Q_{IES}$, $N=3$.}
\label{sim_ch_Q_IES_V1}
\end{figure}

\begin{figure}[b!]
\centering{\includegraphics[width=9cm, trim = 30mm 0mm 20mm 0mm, clip]{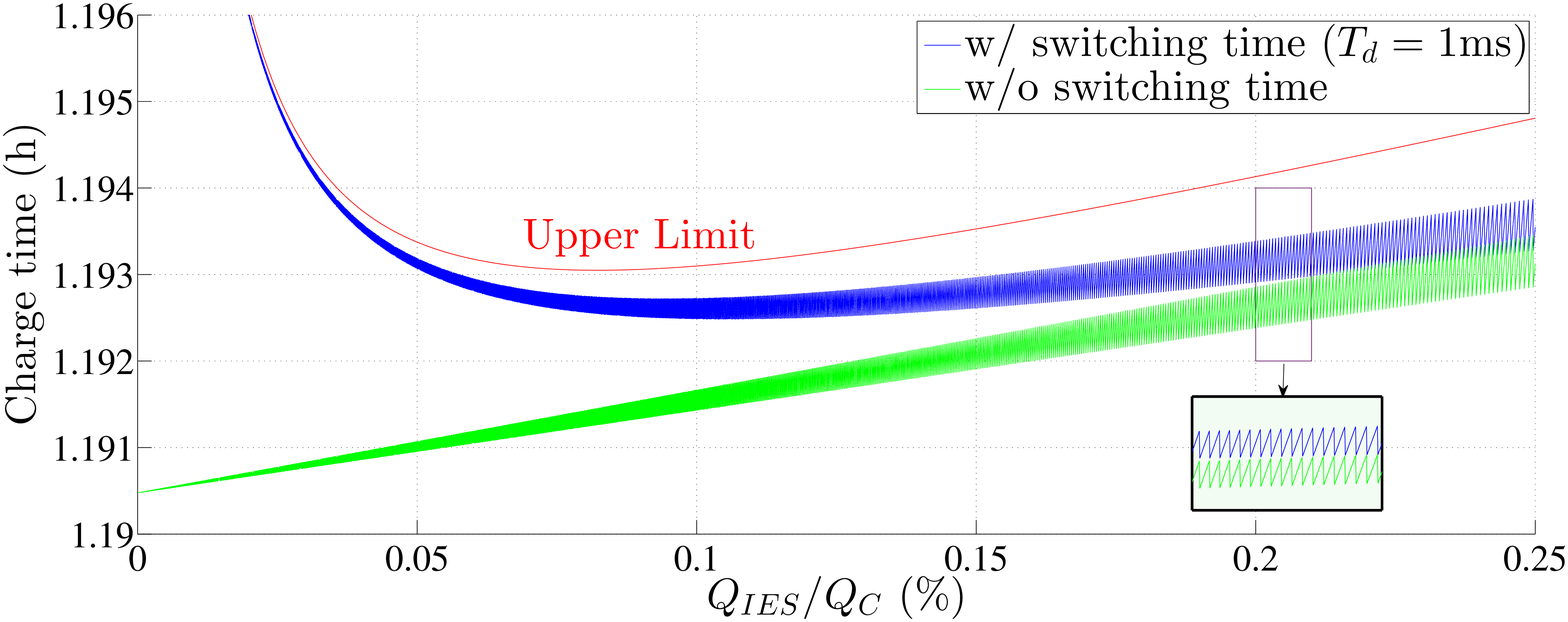}}
\caption{The charging time over $Q_{IES}$, $N=5$.}
\label{sim_ch_Q_IES_V2}
\end{figure}

\begin{figure}[t!]
\centering{\includegraphics[width=8.5cm, trim =30mm 0mm 20mm 0mm, clip]{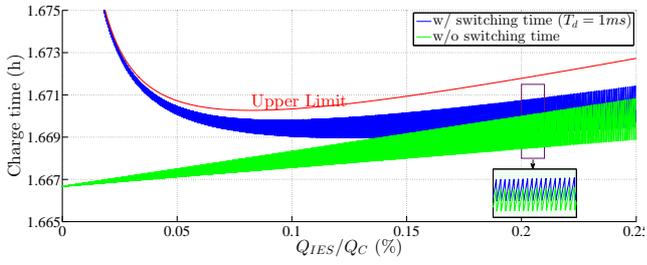}}
\caption{The charging time over $Q_{IES}$, $N=7$.}
\label{sim_ch_Q_IES_V3}
\end{figure}

Figs. \ref{sim_ch_Q_IES_V1},\ref{sim_ch_Q_IES_V2}, and \ref{sim_ch_Q_IES_V3} shows the detailed representation for 2,4, and 6 receivers respectively, and additionally includes the charge time without switch time. Without switching delay, the charge time is reduced as $Q_{IES}$ decreases, while with switching delay, the charge time has a convex waveform over $Q_{IES}$ value. In addition, without the standby time, the charge time lengthens as $Q_{IES}$ increases, while without the standby times, the curve shows Zig-Zag shape over $Q_{IES}$. 
For given $P_R$, $P_B$, $Q_C$ and $N$, the curve increases discretely because the required rounds, $K_L$ is decided discretely over the continuous value of $Q_{IES}$. 
Here, for a single $K_L$ value, the decision equation, in (\ref{T_OC_ST_Td_Full}), has 3 conditions and according to these conditions, $T_{OC}$ increases, decreases and then increases in order as $Q_{IES}$ increases.

\begin{figure}[t!]
\centering{\includegraphics[width= 8.5cm,trim = 40mm 0mm 40mm 0mm, clip]{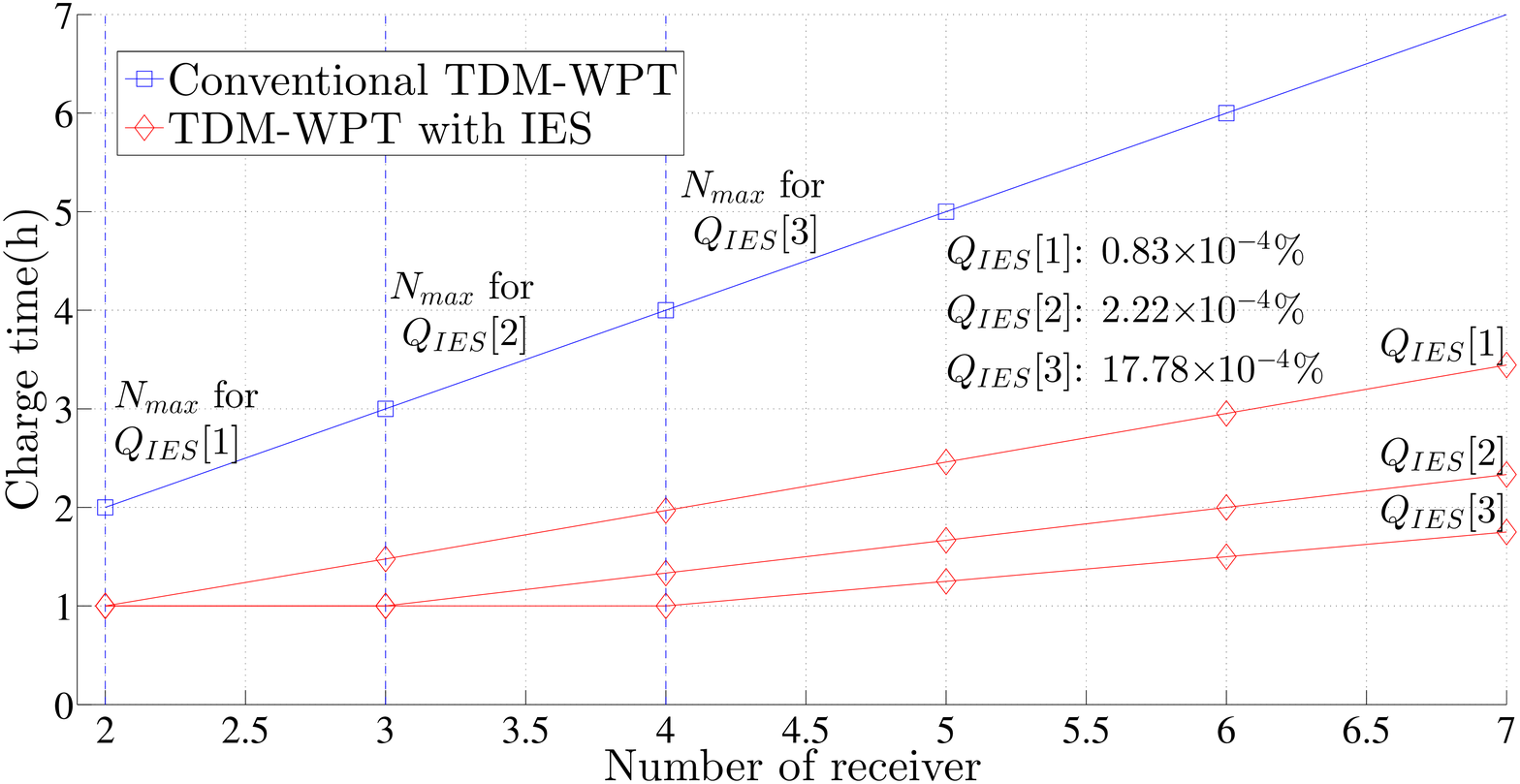}}
\caption{The charge time over the number of receivers.}
\label{sim_ch_time}
\end{figure} 

Fig. \ref{sim_ch_time} shows the upper bound of the charge time over the number of receivers for various capacity ratio of the IES. The selected values of the ratio of $Q_{IES}$ are the optimal values for $N = 2$; $0.83\times10^{-4}\%$, $N = 3$; $2.22\times10^{-4}\%$, and $N = 4$; $17.78\times10^{-4}\%$. 
As $Q_{IES}$ increases, the maximum number of the receivers not to have the standby times, i.e., $N_{max}$, becomes increased to $\left\lfloor P_R/P_B \right\rfloor$, which can be estimated by (\ref{ST_Bound}).
Meanwhile, when $N$ is smaller than $N_{max}$, the total charge time keeps constant value. On the contrary, when $N$ is larger than $N_{max}$, the charge time increases. However it has small increase rate compared to those in the conventional TDM-WPT, which increases to $N$ times for $N$ receivers. Therefore, as the number of receivers increases, the amount of reduced time can be increased further.

\section{Simulation considering practical battery charger}
In addition to the simple simulations for the loads requiring constant power, the simulation considering practical load requirements, which changes continuously, is conducted. In this simulation, the required power for the practical charger depending on the state-of-charge (SOC) of battery \cite{Rahn_2013} is considered.
 
\begin{figure}[b!]
\centering{\includegraphics[width=8.5cm, trim =40mm 0mm 20mm 0mm, clip]{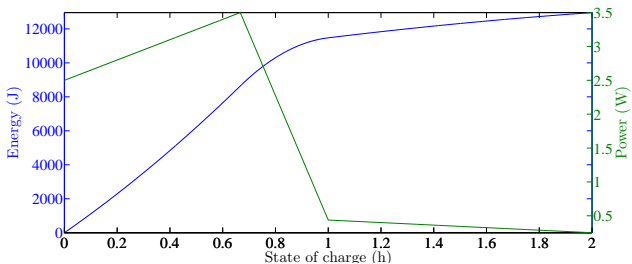}}
\caption{Simplified required power for battery over SOC}
\label{Sim_pw_soc}
\end{figure}

For simplification of practical required power for charger, the battery charing profile $P_B(t)$ is simplified into the piecewise function consisting of 3 linear functions which have different slopes in each interval. Eq.\ref{Bat_SOC} and Fig. \ref{Sim_pw_soc} show the simplified power requirements for SOC which is converted to the time domain.

\begin{equation}
P_B(t) =
\begin{cases}
Q_C(3 + \frac{1}{2000}t), & 0 < t \leq 2400\\
Q_C(\frac{231}{20} - \frac{49}{16000}t), & 2400 < t \leq 3600\\
Q_C(\frac{3}{4} - \frac{1}{16000}t), & 3600 < t \leq 7200
\label{Bat_SOC}
\end{cases}
\end{equation}

\begin{figure}[t!]
\centering{\includegraphics[width=9cm, trim =40mm 0mm 40mm 0mm, clip]{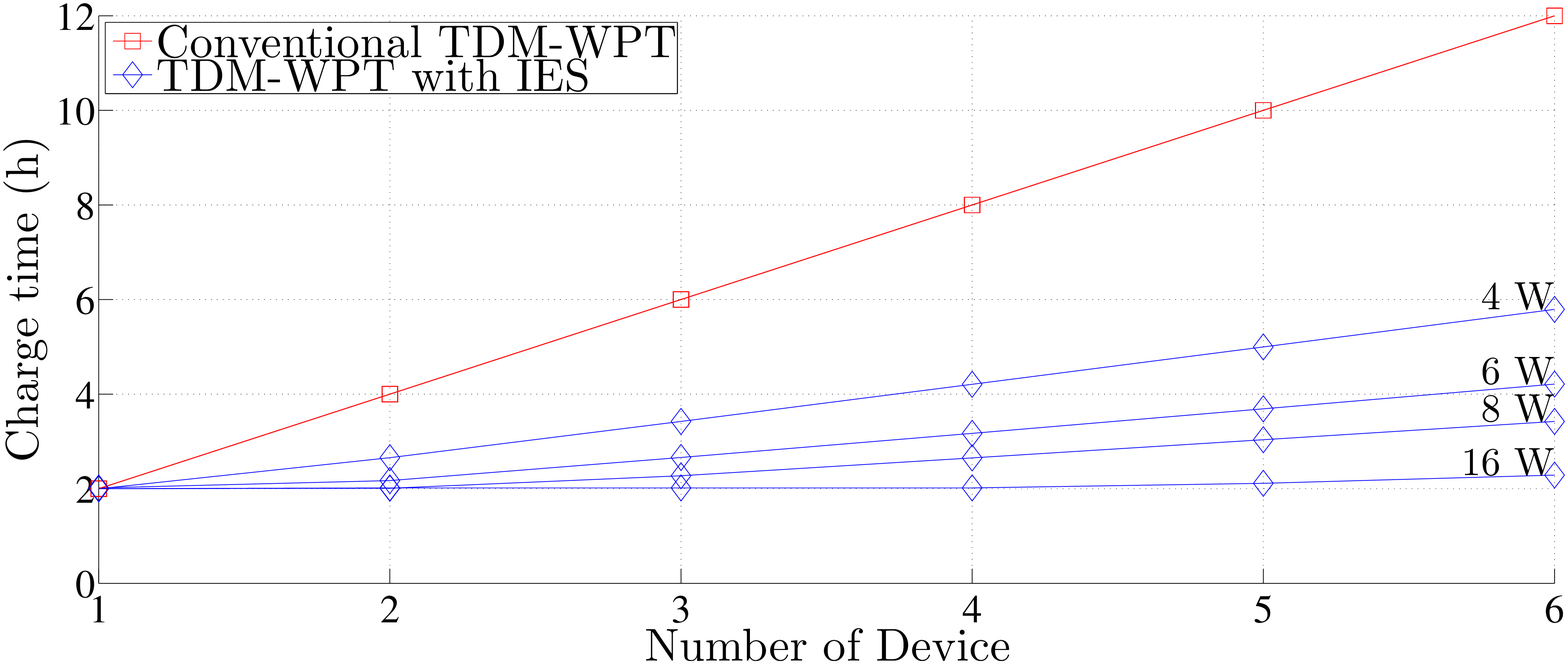}}
\caption{The charge time with the practical charger for various $P_R$}
\label{Sim_Practical_Charge_P_R}
\end{figure}

Fig. \ref{Sim_Practical_Charge_P_R} shows that the charge time can be reduced for the practical charger environments as well. As $P_R$ increases, the reduction of the charge time can also be accelerated. Even in the case that $P_R$ is same as the maximum supplying power of $P_B[t]$, The overall charging time can also be reduced comparing to the conventional one.

\begin{figure}[t!]
\centering{\includegraphics[width=9cm, trim =40mm 0mm 40mm 0mm, clip]{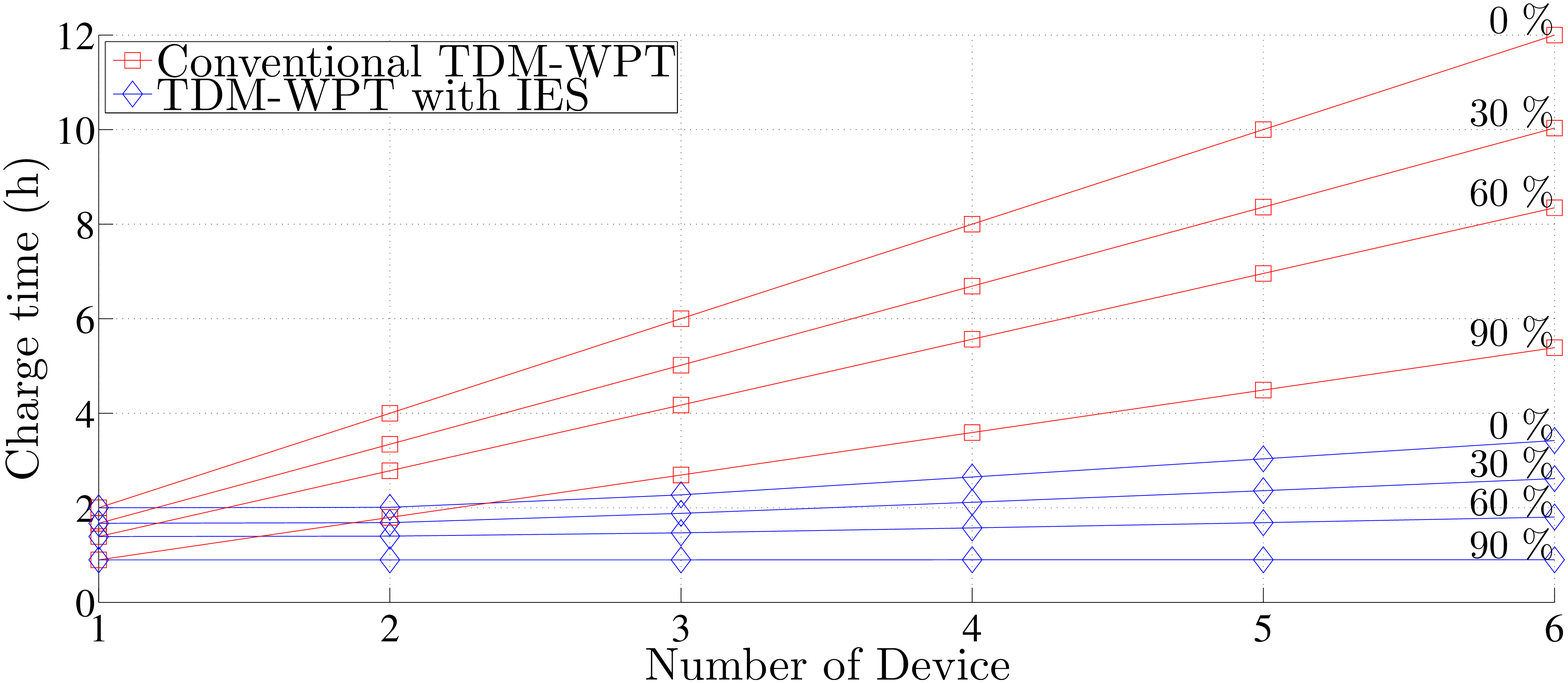}}
\caption{The charge time with the practical charger for various initial SOC}
\label{Sim_Practical_Charge_Init}
\end{figure}

Fig. \ref{Sim_Practical_Charge_Init} shows the effects of the initial SOC of the battery at start-up. The charge time for the proposed system linearly decreases as the initial SOC increases because the IES helps the receiver to absorb the constant power while the power required for load changes according to the SOC.
In the other hand, for the conventional scheme, the changes for the required power explicitly effect on the charge time.
Thus, when charging the battery which has the initial SOC of 60$\%$, it takes longer time because small power is delivered.    
As a result, the proposed scheme provides an additional benefit for easy prediction of the charge time even while the initial SOC changes.

\section{Conclusion}We propose a new multi-device WPT scheme using IES that the total required charge time can be reduced compared to the conventional TDM-WPT. 
By employing the IES, the batteries of different receivers can be charged simultaneously and it eventually leads the charge time reduction. 
In addition, for example of the practical implementation, we present the simple guidelines of some selected key design parameters such as the optimal capacity of IES and the proper number of receivers.
Note that this scheme has far reaching impact because it can also be applied to various power transfer technologies (microwave, laser, etc.) especially based on single transmitter and multiple receivers

\section{Acknowledgment}
This work was supported by the ITRC, Korea.\\


\begin{thebibliography}{1}
\newcommand{\enquote}[1]{``#1''}
\providecommand{\url}[1]{\texttt{#1}}
\providecommand{\urlprefix}{URL }

\bibitem{WPT_2011}
Consortium, W.~P.: , `System description wireless power transfer, volume i: Low
  power, part 1:interface definition, version 1.0.2', , April, 2011

\bibitem{Boys_2000}
Boys, J.~T., Covic, G.~A. Green, A.~W.: `Stability and control of inductively
  coupled power transfer systems', \textit{IEE Proceedings: Electric Power
  Applications}, 2000, \textbf{147}, 1, pp. 37--42

\bibitem{Kurs_2007}
Kurs, A., Karalis, A., Moffatt, R., Joannopoulos, J.~D., Fisher, P. Soljacic,
  M.: `Wireless power transfer via strongly coupled magnetic resonances',
  \textit{Science}, 2007, \textbf{317}, 5834, pp. 83--86

\bibitem{Qualcomm_2010}
Dunworth, J.~D., Martin, R.~W., Selby, M., Maldonado, D., Grilli, F., Velasco,
  J.~T., El-Maleh, K.~H. Karmi, Y.: , `Selective wireless power transfer', ,
  August~4 2010, uS Patent App. 12/850,542

\bibitem{Casa_2009}
Casanova, J.~J., Low, Z.~N. Lin, J.~S.: `A loosely coupled planar wireless
  power system for multiple receivers', \textit{IEEE Trans. Ind. Electron},
  2009, \textbf{56}, 8, pp. 3060--3068

\bibitem{Rahn_2013}
Rahn,C., and Wang,C.~Y.: 'Battery system engineering' (Wiley, Chichester, U.K., 2013) 
\end{thebibliography}
\end{document}